\documentclass[epj,twocolumn]{webofc}
\usepackage[varg]{txfonts}

\def\simless{\mathbin{\lower 3pt\hbox
{$\rlap{\raise 5pt\hbox{$\char'074$}}\mathchar"7218$}}}   %< or of order
\def\simmore{\mathbin{\lower 3pt\hbox
{$\rlap{\raise 5pt\hbox{$\char'076$}}\mathchar"7218$}}}   %> or of order

\woctitle{The Innermost Regions of Relativistic Jets and Their
  Magnetic Fields}
\begin{document}
\title{Numerical study of broadband spectra caused by internal shocks in magnetized relativistic jets of blazars.}
\author{Jes\'{u}s M. Rueda-Becerril\inst{1}\fnsep\thanks{\email{
      jesus.rueda@uv.es}} \and 
  Petar Mimica\inst{1} \and
  Miguel A. Aloy\inst{1} \and
  Carmen Aloy\inst{1}}

\institute{Departamento de Astronom\'{\i}a y Astrof\'{\i}sica,
  Universidad de Valencia, SPAIN}

\abstract{%
  The internal-shocks scenario in relativistic jets has been used to
  explain the variability of blazars' outflow emission. Recent
  simulations have shown that the magnetic field alters the dynamics
  of these shocks producing a whole zoo of spectral energy density
  patterns. However, the role played by magnetization in such
  high-energy emission is still not entirely understood.  With the aid
  of \emph{Fermi}'s second LAT AGN catalog, a comparison with
  observations in the $\gamma$-ray band was performed, in order to
  identify the effects of the magnetic field.}
\maketitle
\section{Introduction}
\label{sec:intro}

Relativistic outflows have been observed extensively in blazars, a
class of radio-loud active galactic nuclei (AGNs) whose jets are
pointing very close to the line of sight towards the observer
\cite{Urry:1995aa}, and known for showing the most rapid variability
of all AGNs. Their remarkable characteristic flares in the X-ray
frequency range usually have a duration of the order of one day. Often
the internal-shocks (IS) scenario \cite{Rees:1994ca} is invoked to
explain this variability
\cite{Spada:2001do,Mimica:2004ay,Mimica:2005aa}. The IS scenario is an
idealized model of a variable jet where an intermittently working
central engine ejects shells of magnetized plasma which collide due to
their velocity differences. As a consequence of the collision internal
shocks are formed, particles are accelerated at the shock fronts and
the non-thermal, highly variable radiation is produced.

Our long-term project is the study of the influence of magnetic fields
on the radiation from IS using numerical simulations. In
\cite{Mimica:2012aa} we studied a large number of shell collisions
with different magnetization levels. In the present work we focus on a
limited number of shell magnetization levels, but vary other
parameters such as the jet viewing angle, bulk Lorentz factor of the
shells, and their relative Lorentz factor. The data obtained from
these simulations is used to categorize the specific effects that
variations of each parameter have on average spectra. These synthetic
observations are then compared with the second LAT AGN catalog (2LAC)
of blazars observed by Fermi \cite{Ackermann:2011apj}.

Our numerical setup is describen in Sec.~\ref{sec:numer-setup}. The
results are shown in Sec.~\ref{sec:results}. Finally we discuss
briefly our results in Sec.~\ref{sec:conclusions}.

\section{Numerical Setup}
\label{sec:numer-setup}

We use a modified version of the \emph{SPEV}
code\cite{Mimica:2009aa,Mimica:2012aa} to compute the non-thermal
emission from the IS. We do not consider the full hydrodynamic
interaction of colliding magnetized shells
(see. e.g. \cite{Mimica:2007aa} for a detailed study). Instead, we
simplify the shell interaction as a one-dimensional Riemann problem
and focus our resources on a more detailed treatment of the
non-thermal radiation. Our method consists of three phases:

\begin{enumerate}
\item \emph{Solution of the Riemann problem.} Making use of an exact RMHD Riemann
  solver \cite{Romero:2005zr} we determine the properties of the
  internal shock waves. We follow the procedure described in
  \cite{Mimica:2010aa} to set-up the shells and to extract the
  information needed for the steps 2 and 3.

\item \emph{Non-thermal particles transport and evolution.} The
  particles are injected behind the shock fronts following the
  prescription of
  \cite{Bottcher:2010gn,Joshi:2011bp,Mimica:2012aa}. We assume that a
  fraction of the thermal electrons are accelerated to high energies,
  and that their energy density is a fraction of the internal energy
  density of the shocked fluid. We assume a cylindrical shell geometry
  and perform all the calculations in the rest frame of the shocked
  fluid. In this frame the shocks are propagating away from the
  initial discontinuity, injecting and leaving non-thermal particles
  behind. We evolve the energy distribution of non-thermal electrons
  taking into account synchrotron and inverse-Compton (IC) losses. See
  \cite{Mimica:2012aa} for more details.

\item \emph{Radiative transfer.} The total emissivity at each point is
  assumed to be a combination of the following emission processes: (1)
  synchrotron radiation, (2) IC upscattering of an external radiation field (EIC)
  and (3) synchrotron self-Compton (SSC). The details of how they are
  calculated are given in \cite{Mimica:2012aa}. The radiative transfer
  equation is solved taking into account the relativistic effects and
  time delays.
\end{enumerate}

We compute light curves and average spectral energy distribution (SED)
for each shell collision. In this work we focus our attention on how
parameter variations affect the SEDs.

\section{Results}
\label{sec:results}

As mentioned in Sec.~\ref{sec:intro}, the aim of this work is to cover
a wider range in the parameter space than was done in
\cite{Mimica:2012aa}. We group our models according to the initial
shell magnetization, $\sigma := B^{2} / 4\pi \rho \Gamma^2 c^2$.
We denote by letters {\bf S}, {\bf M} and {\bf W} the following
families of models:

\begin{itemize}
\item[{\bf W}:] ~~~weakly magnetized, $\sigma_L = 10^{-6}, \sigma_R =
  10^{-6}$,
\item[{\bf M}:] ~~~moderately magnetized, $\sigma_L = 10^{-2},
  \sigma_R = 10^{-2}$, and
\item[ {\bf S}:] ~~~strongly magnetized, $\sigma_L = 1, \sigma_R =
  10^{-1}$.
\end{itemize}

Hereafter the subscripts $L$ and $R$ will denote left (faster) and
right (slower) shells, respectively.  As parameters to vary we
considered both intrinsic and extrinsic ones. Among the intrinsic
parameters we choose the Lorentz factor of the slow shell, $\Gamma_R$,
and the relative Lorentz factor $\Delta g := \Gamma_L/\Gamma_R - 1$,
where $\Gamma_L$ is the Lorentz factor of the fast shell.  The
parameter space covered is shown in Table~\ref{tab:params}.

\begin{table}
  \begin{center}
    \begin{tabular}{lll}
      \hline \hline
      Parameter & value \\
      \hline
      $\sigma_L$ & $10^{-6},\ 10^{-2},\ 1$ \\
      $\sigma_R$ & $10^{-6},\ 10^{-2},\ 10^{-1}$ \\
      $\Gamma_R$ & $10, 12, 17, 20, 22, 25$ \\
      $\Delta g$ & $0.5, 0.7, 1.0, 1.5, 2.0$ \\
      $\theta$ & $5$ \\
      \hline
    \end{tabular}
  \end{center}
  \caption{Parameters of the models. $\Gamma_R$ is the Lorentz factor
    of the slow shell, $\sigma_L$ and $\sigma_R$ are the fast and slow
    shell magnetizations and $\theta$ is viewing angle of the observer.}
  \label{tab:params}
\end{table}

For clarity, when we refer to a particular model we label it by
appending the values of each of these parameters to the model
letter. For instance, {\bf S}-{\bf G}10-{\bf D}1.0-{\bf T}5 is the
strongly magnetized model with $\Gamma_R=10$ ({\bf G}10), $\Delta g =
1.0$ ({\bf D}1.0) and $\theta = 5^\circ$ ({\bf T}5). If we refer to a
subset of models with one or two parameters fixed we use an
abbreviated notation, where we omit the varying parameters from the
label. We compute the spectra for a typical source located at $z=0.5$.

In the rest of this section we will present some of the final SEDs
resulting from our simulations. A larger collection is shown in
\citep{Rueda:2013aa}. The SEDs of each model has been averaged over
the time interval $0-10^{6}$ s.

\subsection{Weakly-magnetized models}
\label{sec:weak}

The SEDs computed for the models {\bf W}-{\bf G}10-{\bf T}5 (varying
$\Delta g$) are shown in the left panel of
Fig.~\ref{fig:W-G10-T5}. The spectra show that with increasing $\Delta
g$ the IC component also increases, up to three orders of
magnitude. In order to see the effects on each emission process, the
synchrotron, SSC and EC components for $\Delta g = 0.5, 2.0$ are shown
as dashed, dot-dashed and dot-dot-dashed lines, respectively. As we
can see, while the three components of the spectrum (synchrotron, SSC
and EC) are around the same order of magnitude for $\Delta g =0.5$,
for $\Delta g = 2.0$ the SSC is almost two orders of magnitude more
luminous than the other two. The inset shows the $\gamma$-ray spectral
slope of each model as a function of its $\gamma$-ray flux (see
Sec.~\ref{sec:spectral-slope}).

\subsection{Moderately-magnetized models}
\label{sec:moderate}

\begin{figure*}
  \centering
  \includegraphics[width=8cm,clip]{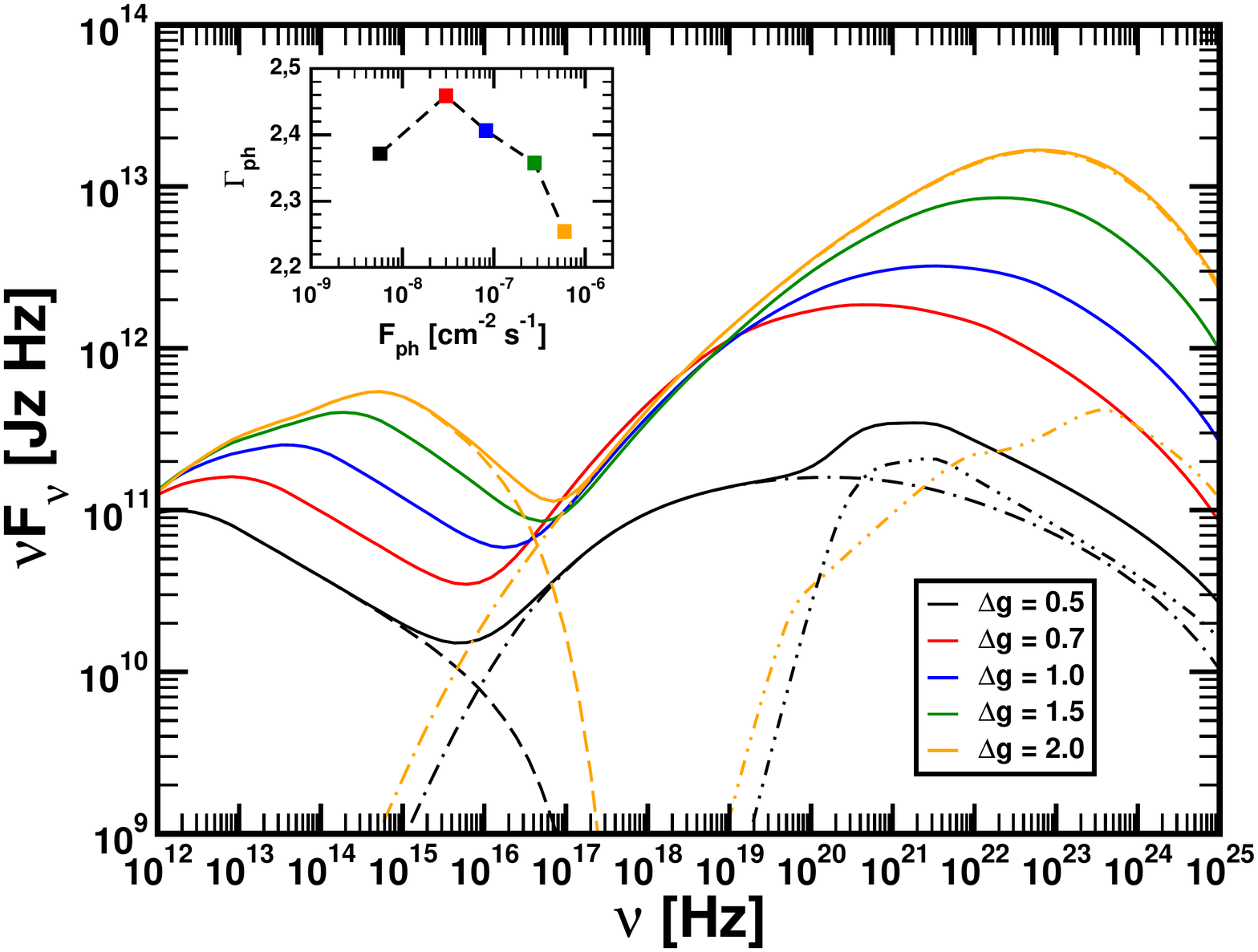}
  \ \hspace{0.16cm}
  \includegraphics[width=8cm,clip]{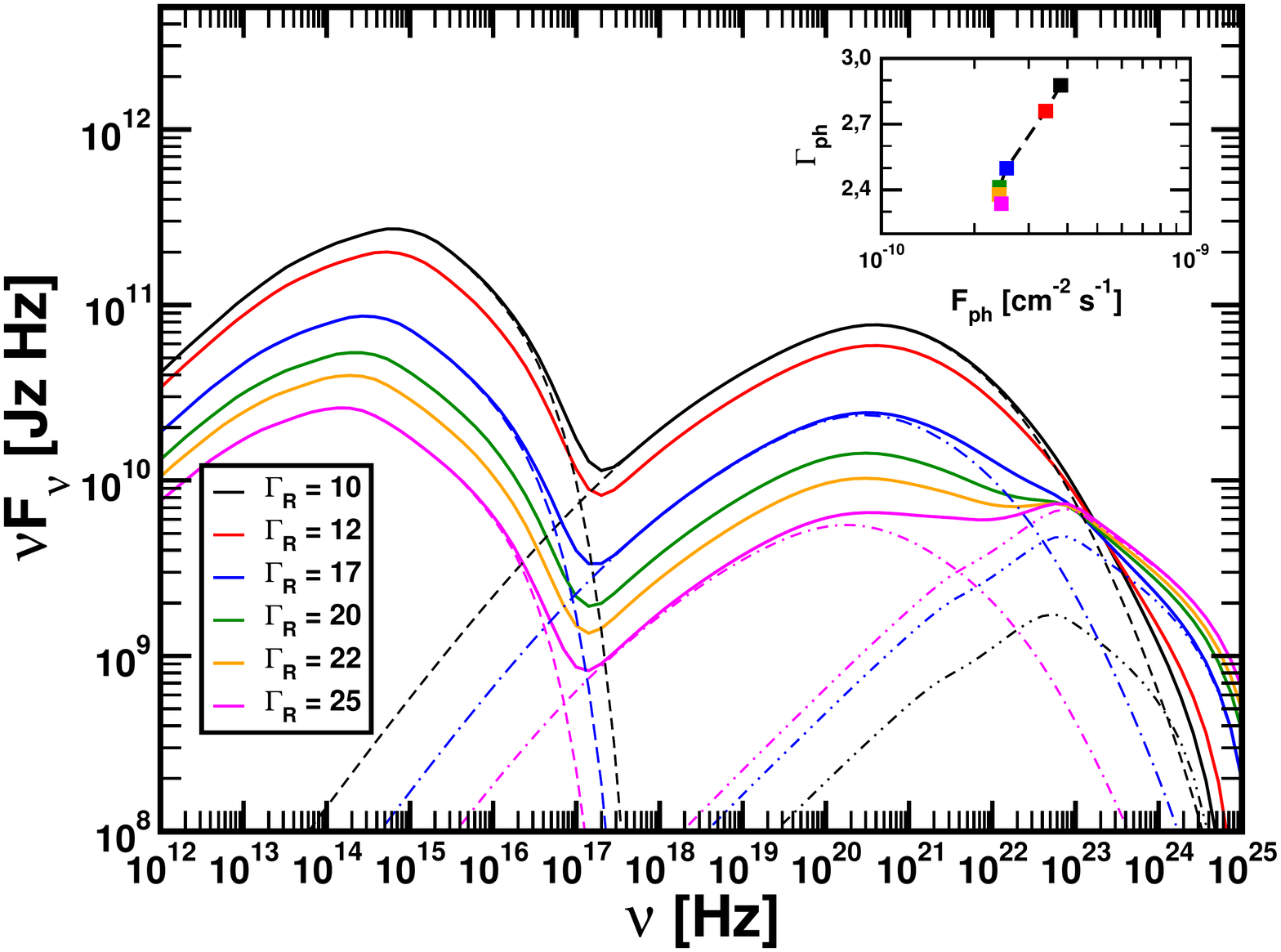}
  \caption{Left panel: Averaged spectra resulting from the collision
    of weakly magnetized shells ($\sigma_L = \sigma_R = 10^{-6}$). For
    the models $\Delta g = 0.5, 2.0$, the synchrotron, SSC and EIC
    contributions (dashed, dot-dashed and dot-dot-dashed lines,
    respectively) are shown. The inset shows the photon spectral slope
    $\Gamma_{ph}$ as a function of the photon flux $F_{ph}$ in the
    $\gamma$-ray band (see Sec.~\ref{sec:spectral-slope}). Colors of
    the points correspond to the line colors in the main plot. Right
    panel: Same as the left panel, but for moderately magnetized
    shells ($\sigma_L = \sigma_R = 10^{-2}$) and varying
    $\Gamma_{R}$. For models with $\Gamma_{R} = 10, 17, 25$ we plot
    the synchrotron, SSC and EIC contributions.}
  \label{fig:W-G10-T5}
  \label{fig:M-D1p0-T5}
\end{figure*}

The SEDs of the family of models {\bf M}-{\bf D}1.0-{\bf T}5 are
presented in the right panel of Fig.~\ref{fig:M-D1p0-T5}. Analogous to
the left panel of Fig.~\ref{fig:W-G10-T5} the synchrotron, SSC and EC
components are shown as dashed, dot-dashed and dot-dot-dashed lines,
respectively, for $\Gamma_R = 10, 17, 25$. The synchrotron component
for $\Gamma_R = 10$ is $\simeq 20$ times brighter than the SSC one, in
contrast to the EC which is $100$ times dimmer. For $\Gamma_R = 25$
the EC is of the same order of magnitude of SSC and synchrotron. The
latter two decrease one order of magnitude between {\bf M}-{\bf
  G}10-{\bf D}1.0-{\bf T}5 and {\bf M}-{\bf G}25-{\bf D}1.0-{\bf T}5,
while the EC grows by almost one order of magnitude. This is a
consequence of the fact that the number of electrons and the comoving
magnetic field strength decrease with the increasing $\Gamma_R$
\cite{Mimica:2012aa}, which means that there are less synchrotron
photons and less electrons which can scatter them in the SSC
process. On the other hand, the radiation field density of the seed
photons for the EC is independent of $\Gamma_R$, which, in combination
with the Doppler boost causes the increase in the EC luminosity. We
also see that at $\simeq 10^{23}$ Hz there is a point where all the EC
spectra coincide. This is due to the Klein-Nishina cutoff, which we
model as a sharp cutoff. The inset shows that the there is no
significant change in the flux of $\gamma$-ray photons, although there
was for the spectral index, heading towards lower values for
increasing $\Gamma_R$.

\subsection{Strongly-magnetized models}
\label{sec:strong}

\begin{figure}
  \centering
  \includegraphics[width=8cm,clip]{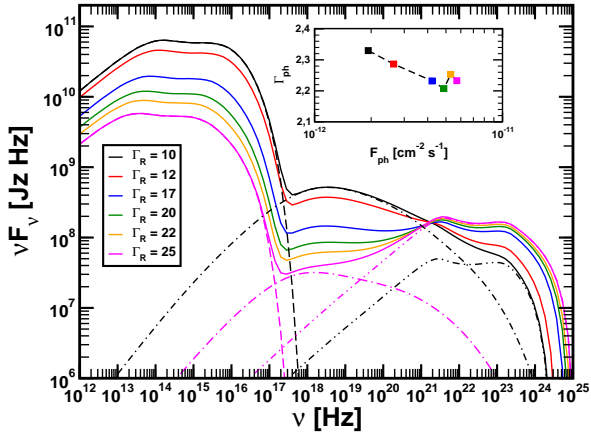}
  \caption{Same as right panel in Fig.~\ref{fig:M-D1p0-T5}, but for
    strongly magnetized shells ($\sigma_L = 1, \sigma_R =$ 0.1).}
  \label{fig:S-D1p0-T5}
\end{figure}

The third model family consists of the strongly magnetized models
where $\sigma_L = 1$ and $\sigma_R = 0.1$. The SEDs of the series of
models {\bf S}-{\bf D}1.0-{\bf T}5 appear in
Fig.~\ref{fig:S-D1p0-T5}. As we can see, for $\Gamma_R = 10$ the
synchrotron component is $\simeq 100$ times brighter than the IC. For
$\Gamma_R = 25$ this difference decreases to one order of
magnitude. The EIC component rises with rising $\Gamma_R$, to the
point in which it begins to be comparable to the synchrotron
component. These effects are similar to the family {\bf M}-{\bf
  D}1.0-{\bf T}5, described in Sec.~\ref{sec:moderate}. The spectra
converge due to our treatment of the Klein-Nishina cutoff. In the
inset we can see that the flux of $\gamma$-ray photons does not change
appreciably in this family of models.

\subsection{$\gamma$-rays spectral slope}
\label{sec:spectral-slope}

\begin{figure}
  \centering
  \includegraphics[width=8cm,clip]{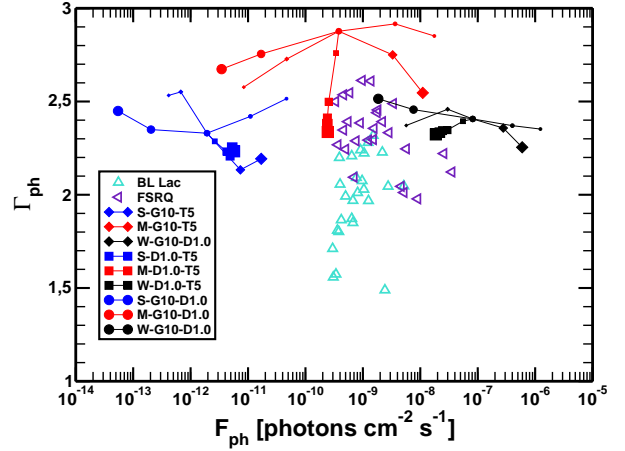}
  \caption{Spectral slope $\Gamma_{ph}$ as a function of the photon
    flux for energies $>100$ MeV. $\Gamma_{ph}$ is computed for the
    photon energies $>200$ MeV \cite{Abdo:2009cb}. The symbols joined
    by lines represent our numerical models, while cyan and magenta
    triangles represent BL Lacs and FSRQs at redshift $z\simeq 0.5$
    from 2LAC \cite{Ackermann:2011apj}. In this figure we also show
    the preliminary three families of models where we vary the opening
    angle (filled circles).}
  \label{fig:spectral-slope}
\end{figure}

A linear mean-squares algorithm is used to deduce the $\gamma$-ray
spectral slope $\Gamma_{ph}$. Due to the fact that we are not modeling
the Klein-Nishina part of the spectrum, we only performed the
calculations of $\Gamma_{ph}$ for those models that do not show a
large drop-off in the photon flux. In Fig.~\ref{fig:spectral-slope} we
show $\Gamma_{ph}$ as a function of the photon flux for energies $>
0.2\,$GeV, where $F_{ph}$ is the photon flux for photon energies $>
0.1$ GeV \cite{Abdo:2009cb}. We compare our results with sources found
in 2LAC catalogue \cite{Ackermann:2011apj} (restricting the comparsion
to sources with $0.4 \leq z \leq 0.6$). In
Fig.~\ref{fig:spectral-slope} we see that weakly and moderately
magnetized models overlap with the observations, with more weakly than
moderately magnetized models falling within the observed part of the
parameter space.

Preliminary results of models where the viewing angle, $\theta$, is
changed; i. e. {\bf SMW}-{\bf G}10-{\bf D}1.0, appear also in
Fig.~\ref{fig:spectral-slope}.

\section{Conclusions}
\label{sec:conclusions}

In this paper we report on the progress of the study of the influence
of the jet magnetization on blazar flares. We vary two parameters of
our models: the relative Lorentz factor $\Delta g$ and the bulk
Lorentz factor $\Gamma_R$.

When $\Delta g$ is increased we get a more luminous maximum of the
inverse Compton component, which is dominated by the SSC. If
$\Gamma_R$ is increased we find that the EIC begins to dominate over
SSC, as well as becoming comparable to the synchrotron component. In
the case of strongly magnetized shells, if $\Gamma_R \sim 50$ both
synchrotron and EC components are of the same order of magnitude (see
\cite{Rueda:2013aa}). Among all the models studied here, the weakly
magnetized are the ones that best fit \emph{Fermi} observations
\cite{Ackermann:2011apj}. However, the tendencies of certain models
with higher magnetization appear to also be consistent with the
observations.

\section*{Acknowledgments}
JMRB acknowledges the support from the Grisolia fellowship
GRISOLIA/2011/041. PM, MAA and CA acknowledge the support from the
ERC grant CAMAP-259276 and the grants AYA2010-21097-C03-01 and
PROMETEO-2009-103.

\bibliographystyle{woc}
\bibliography{RuedaBecerril}

\end{document}